\documentstyle{aipproc}
\input psfig
\begin{document}


\title{Diffractive Phenomena at Tevatron    
\footnote{Published Proceedings from VI Conference on the Intersections of 
 Particle and Nuclear physics, Big Sky, Montana, USA, May 27--June 2 1997.}}


\vspace*{-1.5cm}
\maketitle

\begin{center}A. Santoro\footnote{santoro@lafex.cbpf.br}\\
{\small  For the D\O\ Collaboration\\   
LAFEX/CBPF\\ 
Rio de Janeiro, RJ, Brazil\\
Fermilab\\ P.O.Box 500, Batavia, Il\\60510, USA}
\end{center}

\vspace*{-1cm}
\begin{abstract}

Preliminary results from the D\O\ experiment on jet production with rapidity 
gaps in $p\overline{p}$ collisions are presented.  A class of dijet events with a 
forward rapidity gap is observed at center-of-mass energies $\sqrt{s}$ = 1800 
GeV and 630 GeV.   The number of events with rapidity gaps at both 
center-of-mass energies is significantly greater than the expectation from
multiplicity fluctuations and is consistent with a hard single diffractive 
process.  A class of events with two forward gaps and central dijets are 
also observed at 1800 GeV. This topology is consistent with hard double 
pomeron exchange. We also present proposed plans for extending these 
analysis into Run II through the use of a forward proton detector.
	
\end{abstract}


\section{Introduction}

	The results coming from Tevatron (CDF and D\O)
and Hera (H1 and ZEUS) \cite{agb:TEVATRON,agb:D0PP96} and the progress in the phenomenological
and theoretical side of diffractive physics are very important to
understand the complexity of the pomeron in high energy physics.  
	Soft and hard diffraction are an important part of strong interactions.
Soft diffraction is well described phenomenologically by the Regge Model
(RM) with a hadronic pomeron \cite{agb:ALBERI}. For hard diffraction 
\cite{agb:INGELMAN}
we have many interesting models proposing different configurations for the 
pomeron as a composite object. 
Among the leading models for a hard pomeron we have the ``economical'' 
two-gluon model \cite{agb:NUSSINOV} and the ``hot spot of gluons'' or the BFKL
pomeron \cite{agb:BFKL}.

   Hard diffraction can be best studied at the Tevatron due to the large 
diffractive mass accessible. For hard single diffraction  we have aproximately
$\rm{M_{x}}$=400 GeV and for double pomeron exchange  $\rm{M_{x}}$=100 GeV, 
where $\rm{M_{x}}$ is the diffractive mass produced by interactions with pomeron.
	We believe that these subjects are certainly a good bridge between the
soft (RM) and  hard (QCD) \cite{agb:INGELMAN} interactions.
	We will present here some preliminary results from D\O\
for hard single diffraction and for double pomeron exchange \cite{agb:D0PP96}.

\section{hard Single diffraction}
	An experimental signature of hard diffractive events is the presence
of a rapidity gap (lack of particle production in a rapidity or 
pseudorapidity\footnote{Pseudorapidity or $\eta=-ln[tan(\frac{\theta}{2})]$,
where $\theta$ is the polar angle defined relative to the proton beam 
direction.} region), along with a hard scattering (jet production, 
$W$ production, etc.). Since the pomeron is a color singlet, radiation is 
suppressed in events with pomeron exchange, typically resulting in a large 
rapidity gap~\cite{agb:DIETER}. In hard single diffraction a pomeron is emitted
from one of the incident particles (proton or antiproton) and undergoes a 
hard scattering with the second proton, often leaving a rapidity gap in the 
direction of its parent particle (antiproton or proton). We examine the 
process $p + \overline{p} \rightarrow j + j + X$ and look for the presence 
of a forward rapidity gap along the direction of one of the initial beam 
particles.  

The existence of a diffractive signal in the experimental data may be observed 
as a larger number of rapidity gap events in the forward multiplicity 
distribution than expected from the non-diffractive background.
Given sufficient detector resolution, sensitivity, and statistics, 
two components in the multiplicity distribution can be resolved and 
the relative fraction of rapidity gap events in excess of expectations from 
a smoothly falling multiplicity distribution can be estimated.

The D\O\ detector~\cite{agb:NIM} is used to provide experimental information on 
the fraction of jet events with forward rapidity gaps. This analysis primarily 
utilizes the uranium-liquid argon calorimeters which have full coverage for a
pseudorapidity range of $|\eta|\!<\!4.1$. The transverse segmentation of the 
projective calorimeter towers is typically $\Delta\eta \mbox{$\, \times \,$} 
\Delta\phi = 0.1 \mbox{$\, \times \,$} 0.1$. The electromagnetic (EM) section 
of the calorimeters is used to search for rapidity gaps.  The EM section is 
particularly useful for identifying low energy particles due to its low level 
of noise and ability to detect neutral pions.  A particle is tagged by the 
deposition of more than 200\,MeV of energy in a single EM calorimeter tower.

\begin{figure}[t]
\vspace{.5cm}
\vbox{
\centerline{\psfig{figure=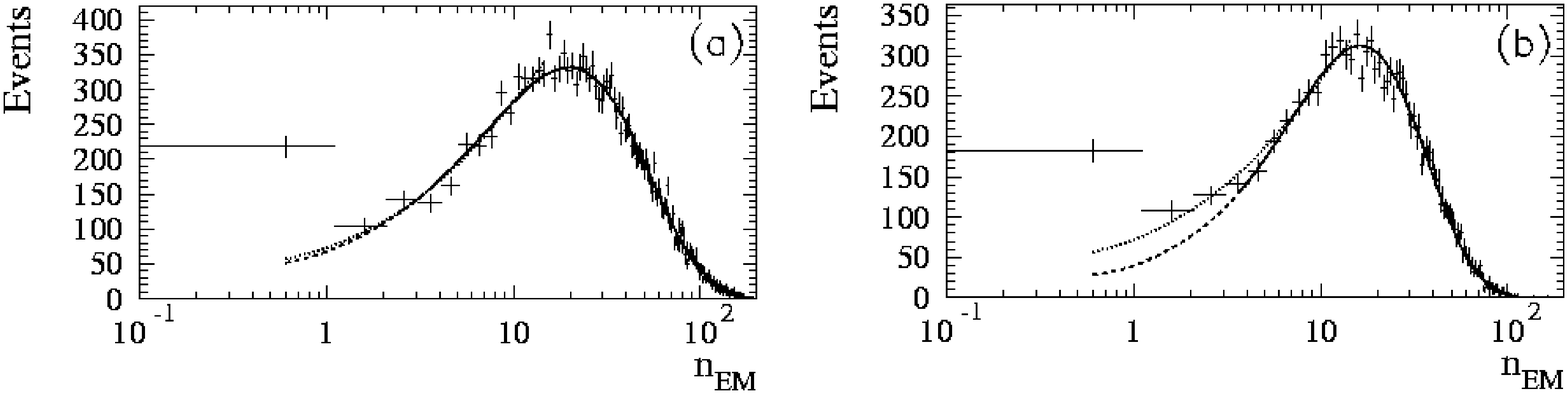,height=5cm,width=14.cm}}
\caption{Number of electromagnetic calorimeter 
towers ($\rm{n_{EM}}$) above a $200$\,MeV energy threshold for the region
$2<\eta<4.1$ opposite the forward jets for center-of-mass energies
of (a) $1800$\,GeV and (b) $630$\,GeV.
The curves are negative binomial fits to the data 
excluding low multiplicity bins.
}
\label{agb:fwd_mult}
}
\end{figure}

The data used in this study were obtained using a forward trigger requiring 
at least two jets above $12$\,GeV in the same hemisphere with both jets 
having $\eta>1.6$ or $\eta< -1.6$.  Since the pomeron only carries a few per 
cent of the initial proton momentum, the jet system is expected to be boosted 
in diffractive jet production, thus a forward jet trigger
can be utilized to provide an enhanced sample of diffractive events.
Offline, two jets above trigger threshold are required and
events with multiple $p\overline{p}$ interactions or spurious
jets are removed. Jets are reconstructed using 
a cone algorithm with radius, $R = \sqrt{\Delta\eta^2 + \Delta\phi^2}=0.7$.
The number of EM towers ($\rm{n_{EM}}$) above a $200$\,MeV energy threshold 
is measured in the hemisphere opposite the leading two jets in the region 
$2<|\eta|<4.1$.  The ($\rm{n_{EM}}$) distribution for 
the forward trigger is shown in Fig.~\ref{agb:fwd_mult}
for $\sqrt{s}$ of (a) $1800$\,GeV and (b) $630$\,GeV.

\newpage

\vspace*{-1.cm}

The distributions at both center-of-mass energies
show a peak at zero multiplicity in qualitative
agreement with expectations for a diffractive signal component. 
Negative binomial fits to the leading edge and the whole distribution
(excluding $\rm{n_{EM}}=0$)  have been used to estimate the non-diffractive
background.
A fractional excess of rapidity gap events is defined to be the number of
zero multiplicity events in excess of those predicted by the fit
divided by the total number of events in the sample.
The fractional excess observed 
in the forward region for the 
$\sqrt{s} = 1800$\,GeV sample
is $0.67 \pm 0.05 \%$,
where the error includes only statistical uncertainties and a systematic 
uncertainty based on the choice of range for the fit.  
An excess of rapidity gap events is also clearly observed 
at $630$\,GeV with a magnitude of $1-2\%$.  
Systematic studies have not been completed, but effects such as
gap detection efficiency are expected to reduce the number of observed
rapidity gaps, and
correcting for these effects is expected to give a modest increase in
the magnitude of the signal measurement.
The observed fractional excess is relatively
insensitive to the calorimeter energy threshold, and
rapidity gap events ($\rm{n_{EM}}=0$) typically have zero multiplicity
in other available detectors, such as hadronic calorimeters, forward 
tracking, beam hodoscopes, and forward muon chambers.  

The forward gap fraction measurement for the
$\sqrt{s} = 1800$\,GeV sample has
been extended to unrestricted jet
topologies using an inclusive jet trigger
and we observe that the  gap fraction
increases with the boost of the jets, consistent with the expected
behavior of diffractive events discussed earlier.

\section{Double pomeron exchange}

\begin{figure}[ht]
\vspace{.5cm}
\vbox{
\centerline{\psfig{figure=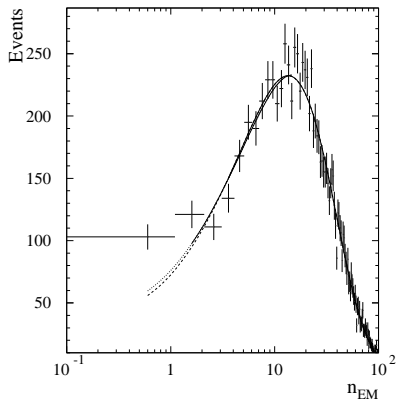,height=6cm,width=6cm}}
\vspace{-1.5cm}\caption{The $\rm{n_{EM}}$ distribution opposite
the tagged gap for single gap trigger data.  The zero multiplicity events
are double gap events in this sample.  The curves
are negative binomial fits to the data excluding low multiplicity bins 
as described in the previous section.}
\label{agb:sv_data}
}
\end{figure}

The same experimental methods may be applied to a search for hard double pomeron
exchange.  In this process both incoming protons emit a pomeron and the two
pomerons interact to produce a jet system.  Rapidity gaps are expected to be
produced along each forward beam direction, since there is no color connection
between the jet system and the beam particles.
In this analysis we have selected an enhanced sample of forward rapidity
gap events with a dedicated single gap trigger, which, in addition to the
jet requirements, vetoes
on forward particles in either beam direction using the scintillator
beam hodoscopes which bracket the D\O\ collision region.
Events were selected to have a rapidity gap ($\rm{n_{EM}}=0$) in 
the direction of the online veto.  These data consist of about 40,000 
single gap events at 
$\sqrt{s}=1800$\,GeV, compared to the approximately 200 events observed
in the forward trigger sample after background subtraction.
This enhanced diffractive sample is used to search for double forward gap
events, in which we require no towers above threshold in both forward
calorimeter regions along with two jets with $E_T>15$\,GeV and 
$|\eta|<1.0$.
This is an expected topology for events produced in hard double pomeron
exchange.  The $\rm{n_{EM}}$ distribution for the veto-trigger
is plotted in Fig.~\ref{agb:sv_data} for the forward region ($2<|\eta|<4.1$)
opposite the tagged rapidity gap.  
We clearly observe a sample of double gap events, although an interpretation
of them in terms of hard double pomeron exchange requires further study.

\section{Conclusion and Prospects For The Next Future -Run II at the Tevatron}

We have observed the presence of forward rapidity gaps in events with high 
$E_T$ jet production with the D\O\ detector at Fermilab.  The fraction of
forward rapidity gap events observed is in excess of those expected to be
produced via multiplicity fluctuations at center-of-mass energies of
$1800$\,GeV and $630$\,GeV. This is consistent with expectations from hard 
single diffractive jet production and provides the first experimental evidence
for this process at $\sqrt{s}=1800$\,GeV.  We also observe a class of events 
containing high $E_T$ central jets and two forward rapidity gaps, consistent
with a hard double pomeron exchange event topology.

	 All these results motivated us to propose a new set of sub-detectors 
(the so called Roman Pots \cite{agb:POTS}) to be introduced in the two arms of the 
D\O\ Detector.  This forward proton detector (FPD) is composed of four quadrupole 
spectrometers and one dipole spectrometer. They will tag the scattered proton 
and/or anti-proton, allowing us to observe directly a number of 
processes like double pomeron exchange and improve our studies of all 
diffractive topics in Run II at the Tevatron.

We acknowledge the support of the US Department of Energy and
      the collaborating institutions and their funding agencies in this
      work.



\end{document}